# Forward Link Interference Mitigation in Mobile Interactive Satellite Systems

P. Henarejos[*], M. A. Vázquez[†] and G. Cocco[‡]
*Centre Tecnològic de Telecomunicacions de Catalunya, 08860 Castelldefels, Barcelona, Spain*

A. Pérez-Neira[§]
*Universitat Politècnica de Catalunya, 08034 Barcelona, Spain*

**We present the results of the performance evaluation of polarization-time coding and soft interference cancellation in multi-beam satellite systems affected by co-channel interference in realistic setups. The standard of Broadband Global Area Network service (BGAN) has been considered as reference for the physical layer and realistic interference and channel models have been adopted. The work has been carried out in the framework of the Next Generation Waveform for Increased Spectral Efficiency (NGWISE) project founded by the European Space Agency (ESA).**

## I. Introduction

Interactive satellite systems are a key communication solution with a huge potential market. Possible applications range from the provision of data connectivity to areas where cellular connection is not profitable (e.g., rural) or infeasible (e.g., maritime and aircraft scenarios) to backing up in emergency situations, especially when involving large geographical areas. Far from being limited to broadcast transmissions, the above mentioned applications rely on multicast and multiple unicast connections.

In this framework multi-beam satellites can provide a many-folds increase in spectral efficiency with respect to global beam satellites, especially in case of architectures with a low frequency reuse factor. An even higher throughput can be provided, in principle, by leverage on polarization reusing on adjacent beams and transmission on both polarizations within a single beam. However, the adoption of aggressive frequency/polarization reuse schemes implies an increase of intra-system interference due to satellite antenna side-lobes, low directivity of user terminal (UT) antennas and polarization mismatch due to antennas imperfections and atmospheric propagation.[1] Moreover, mobile terminals are also keen to propagation impairment such as shadowing and fading, while the large propagation delay typical of satellite systems (especially GEO) prevents the availability of channel state information (CSIT).

Time diversity is largely exploited in today's interactive mobile systems standards to overcome channel impairments in mobile broadcast systems. However, new diversity techniques have recently gained interest. Such techniques are based on polarization and spatial diversity that allow to apply multiple input-multiple output (MIMO) techniques such as precoding and polarization-time codes.

Interference cancellation techniques[2] are also a potential solution that is currently being looked at in both the forward and the reverse link.

Dual polarization transmission has been evaluated for the mobile broadcast scenario in[3] with promising results. The joint effect of outdated CSIT and the time variability of channel makes very difficult the use of linear precoding as it was applied in previous works.[4,5] Apart from an increase in system diversity, the dual polarization transmission can provide an increase in spectral efficiency especially in case of low cross-polar interference.

---
[*]Research Engineer, Communication Systems, pol.henarejos@cttc.es
[†]Research Engineer, Communication Systems, mavazquez@cttc.es
[‡]Post-doctorate Researcher, Communication Systems, gcocco@cttc.es
[§]Professor at Signal Theory and Communications Department at UPC and Senior Researcher at CTTC, ana.isabel.perez@upc.edu

In the present paper present part of the results we obtained within the Next Generation Waveform for Increased Spectral Efficiency (NGWISE) project founded by the European Space Agency (ESA).[6] More specifically, we evaluate the impact of dual polarization transmission in terms of throughput in presence of co-channel interference in a multi-beam satellite system and evaluate the possibility of applying soft interference cancellation (SIC) in interference-limited setups. A realistic channel model is adopted and different scenarios are considered, namely maritime and terrestrial. The standard adopted in Broadband Global Area Network service (BGAN)[7] is used as reference standard for the physical layer. In the rest of the paper we will refer to [7] as BGAN standard for simplicity.

Our results show that that a higher spectral efficiency can be achieved through the considered techniques, which may lead to an increased overall system throughput. We also show that for a four-color frequency reuse scheme soft interference cancellation provides limited gain in terms of block error rate due to the low relative power and high number of interferers that can be assimilated to Gaussian noise.

## II.  System Model

Let us consider the forward link of a multi-beam geostationary satellite communication system. Co-channel interference among adjacent beams is mitigated through frequency reusing. A four colors frequency reuse scheme is considered in the following. A dual polarization transmission is assumed, i.e., the satellite and the user terminal antennas transmits and receive over two (almost-)orthogonal polarizations, respectively. The received signal at the user terminal can be expressed as:

$$\mathbf{Y} = \sqrt{P}\mathbf{HBC}(\mathbf{s}) + \mathbf{HBJ} + \mathbf{N} \qquad (1)$$

where $\mathbf{Y} \in \mathbb{C}^{2\times 2}$ represents the received signal in two time instants from the two polarizations, $P$ is the transmitted power, $\mathbf{H} \in \mathbb{C}^{2\times 2}$ is the channel matrix, the distribution of which depends on the scenario, $\mathbf{B} \in \mathbb{R}^{2\times 2}$ is a matrix that accounts for the antennas characteristics in terms of co-polar, cross-polar gains and co-channel interference rejection while $\mathbf{C}(\mathbf{s})$ is the polarization-time code, which depends on the symbol vector $\mathbf{s} \in \mathbb{C}^{M\times 1}$, having block length $M$. We assume a general complex symbol mapping. The inter-beam interference is modelled through the matrix $\mathbf{J} \in \mathbb{C}^{2\times 2}$. Thermal noise is taken into account through the term $\mathbf{N} \in \mathbb{C}^{2\times 2}$ whose entries are zero mean Gaussian random variables with variance $N$.

As is often the case in terrestrial communications, also in the satellite context the use of MIMO techniques is combined with channel coding leading to what is called a MIMO-BICM modulation. Figure 1 depicts the general block diagram of this scheme. The block $\Pi$ and $\Pi^{-1}$ in the picture represent the coded bit interleaver and deinterleaver, respectively.

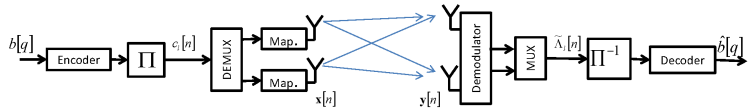

Figure 1.  Equivalent MIMO-BICM scheme of the considered dual polarization setup.

For this scenario, the main challenges to be faced are two. First, the MIMO transmission scheme (i.e. how symbols are transmitted by the two polarizations) must be decided. Secondly, the MIMO demodulator (i.e. how symbols from the two polarizations are detected) needs to be also obtained.

For both the reference signal (i.e., the signal that is to be received by the user terminal) and the interfering signals the BGAN standard[7] is adopted. The BGAN standard, currently under definition, is designed to support both voice and broadband data services in a wide range of scenario such as maritime and land mobile. The channel code adopted in the BGAN standard is a turbo code with several possible configurations in terms of code rate and codeword length. The possible combinations of channel code parameters, modulation (QPSK, 16QAM), symbol rate and other physical layer characteristics are defined by the "bearers".

## III.  Polarization-Time Transmission and Reception Schemes

In the present section we consider several possible MIMO modulation and demodulation solutions previously proposed in literature that are suited for application in the considered setup. Out of them, one modulation and one demodulation scheme are selected based on practical as well as theoretical considerations.

### III.A. MIMO Transmission Scheme

- **Alamouti**: The objective of this polarization-time code is to obtain the maximum diversity gain by constructively adding up the channel gains from the two polarizations while rejecting the inter-symbol interference (ISI).[8] In this case $M = 2$ and the coding matrix is

$$\mathbf{C}_{\text{ala}}(\mathbf{s}) = \begin{pmatrix} s_1 & -s_1^* \\ s_2 & s_2^* \end{pmatrix}. \tag{2}$$

The Maximum-Likelihood (ML) detection for this scheme is known to have low complexity, as it reduces to to a matrix multiplication and a set of comparisons. After optimally decoding this polarization-time code, the symbol obtains a diversity gain of

$$G_d = \|\mathbf{HB}\|^2 = |\mathbf{h}_{11}|^2 b_{11}^2 + |\mathbf{h}_{22}|^2 b_{22}^2, \tag{3}$$

where $b_{11}$ and $b_{22}$ are the diagonal elements of $\mathbf{B}$. Note that this equivalent SISO gain is obtained when the ML detection is used.

- **Polarization Multiplexing**: Polarization multiplexing obtains the full multiplexing gain (i.e. in two time instants 4 symbols are transmitted, 1 for each channel use of each antenna).[9] For this case $M = 4$ and the coding matrix is

$$\mathbf{C}_{\text{mul}}(\mathbf{s}) = \begin{pmatrix} s_1 & s_3 \\ s_2 & s_4 \end{pmatrix}. \tag{4}$$

- **Golden Code**: this is a full diversity technique that still provides some coding gain.[10] The coding matrix is constructed as follows

$$\mathbf{C}_{\text{gol}}(\mathbf{s}) = \begin{pmatrix} s_1 + \alpha s_2 & s_3 + \alpha s_4 \\ i(s_3 + \beta s_4) & s_1 + \beta s_2 \end{pmatrix}, \tag{5}$$

where $\alpha = \frac{1+\sqrt{5}}{2}$ and $\beta = \frac{1-\sqrt{5}}{2}$.

### III.B. MIMO-BICM Demodulators

A low complexity detector for the MIMO schemes presented above is the hard decision ML detector. The ML decision rule consists in solving the following optimization problem:

$$\arg\min_{\hat{\mathbf{s}}} \quad \|\mathbf{y} - \mathbf{HC}(\hat{s})\|_2^2. \tag{6}$$

Note that the detected symbol $\hat{s}$ is obtained via a hard decision. However, a channel code is usually included in all MIMO schemes, and thus the MIMO demodulator should output the log-likelihood ratios (LLR) for the coded symbols that are to be passed to the decoder[11,12]. The LLR for the $l$-th coded symbol is

$$\Lambda_l = \log \frac{p(c_l = 1|\mathbf{y}, \mathbf{H})}{p(c_l = 0|\mathbf{y}, \mathbf{H})}, \tag{7}$$

where $p(c_l|\mathbf{y}, \mathbf{H})$ is the probability mass function of the coded bits conditioned on the channel output $\mathbf{y}$ and the channel matrix $\mathbf{H}$. In order to reduce the complexity the log-sum approximation can be applied with little loss in terms of performance.[11] In the following we consider different schemes that aim at decreasing the demodulator complexity.

In order to reduce the number of demodulators to test, we refer to the recommendations in[13] where several demodulators were studied considering the mutual information as a measurement. In[13] it was shown that for low data rates the soft-minimum mean square error (MMSE)[14,15] demodulator outperforms the other designs. As an extension of this technique we propose the soft version of the MMSE-SIC receiver. In the following briefly describe such schemes as well as the optimal solution for the uncoded case.

- **Soft-MMSE**: this is a linear equalizer obtained through the minimization of the mean square error (MSE). Its expression is

$$\mathbf{G}_{\text{MMSE}} = \left(\mathbf{H}^H \mathbf{H} + \sigma^2 \mathbf{I}\right)^{-1} \mathbf{H}^H. \tag{8}$$

- **Soft-MMSE-SIC**: this detector is the soft version of the V-BLAST demodulator. This technique iteratively decodes and subtracts the ISI. In order to adapt this to our setup we substitute the hard decision of the interference with a soft one. The resulting algorithm is the following:

  1. Find the least faded polarization:
  $$k_i = \min_k g_k \qquad (9)$$
  where $g_k$, $k = 1, 2$ are the components of vector $\mathbf{g} = \text{diag}\left(\left(\mathbf{H}\mathbf{H}^H\right)^{-1}\right)$.

  2. Obtain the MMSE estimate of signal $s_{k_i}$ transmitted in the polarization $k_i$:
  $$\hat{s_{k_i}} = [\mathbf{G}_{\text{MMSE}}\mathbf{y}_n]_{k_i}. \qquad (10)$$

  3. Subtract interference due to this symbol from $\mathbf{y}$
  $$\mathbf{y}_n = \mathbf{y}_n - \hat{s_{k_i}}[\mathbf{H}]_{k_i}, \qquad (11)$$
  where $\mathbf{y}_n$ is the received signal at time instant $n$.

  4. Remove $k_i$th component from $\mathbf{y}_n$ and the $k_i$th column of $\mathbf{H}$.

  5. Repeat 1-4 for the other polarization.

### III.C. System Design

We chose one encoding/decoding scheme to be applied in our scenario among those presented so far. The choice has been a trade off between theoretical considerations and practical constraints either dictated by the BGAN standard or by other practical considerations such as complexity at the user terminal.

- Transmission scheme: polarization multiplexing. It was shown in previous ESA projects that this method outperforms the Alamouti one when considering a given transmit power. Another possible choice would be to use Golden codes that, according to some preliminary results we obtained, shows an improve of around 1dB in FER with respect to Alamouti. However the Golden codes imply an increase in computational complexity as 4 symbols are mixed together in two time instants, which makes the detection more computationally demanding. Keeping in mind that in our study case low complexity at the receiver is an asset, we considered that the enhancement in terms of FER does not justify the increase in complexity.

- Detection scheme: Soft-MMSE-SIC. As described in[13] this scheme outperforms the other demodulators and presents a good performance in the higher spectral efficiency region. As a matter of facts it can be observed in figures 2 and 3 that the soft-MMSE demodulator performs slightly better in low data rates while at higher data rates it can be observed that the Soft-MMSE-SIC performs better.

## IV. Soft Interference Cancelation

The diversity/multiplexing gains potentially delivered by polarization-time schemes may suffer from the co-channel interference coming from other beams depending on the interference strength. In first approximation (more accurate if the number of interferers is large) such interference can be assimilated to a background noise which can not be dealt with using the MIMO techniques presented in Section III. Interference cancelation may help in such case. A first classification of interference cancelation methods can be done by distinguishing hard (HIC) and soft interference cancelation (SIC). In HIC one of the signals (usually the strongest one) is decoded treating the others as noise and than subtracted from the received waveform. Such scheme is relatively simple but has the drawback that, if the signal to be subtracted is not decoded correctly, error propagation can severely limit the performance of the system. SIC methods consist of a soft estimation of each of the transmitted signals followed by a decoding phase in which such estimation is taken into account by the decoder. Examples can be found in,[16][17] and.[18] In the following we consider the iterative SIC scheme depicted in Fig. 4 for the case of two received signals (one reference signal and one interferer) in a SISO channel.

In such scheme the received waveform is fed to the soft estimator, which performs detection and estimates the transmitted channel symbols for both signals. The estimation is performed using a turbo decoder that

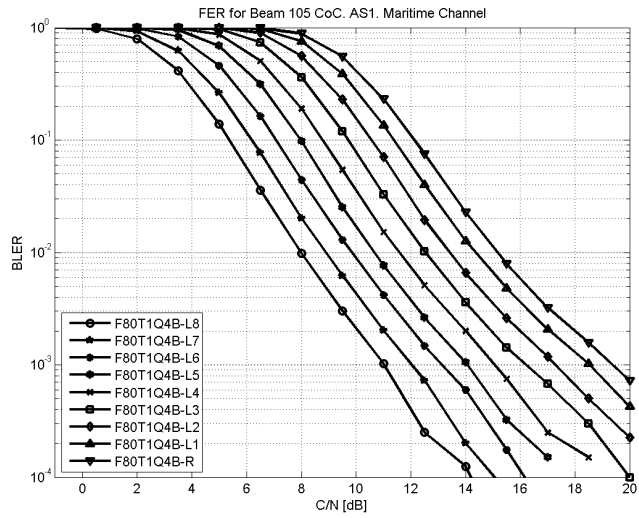

Figure 2. BLER for Soft-MMSE receiver at center of coverage for maritime scenario.

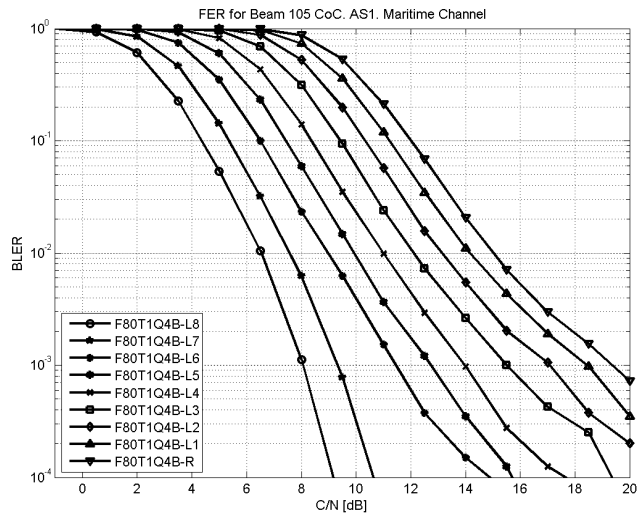

Figure 3. BLER for Soft-MMSE-SIC receiver at center of coverage for maritime scenario.

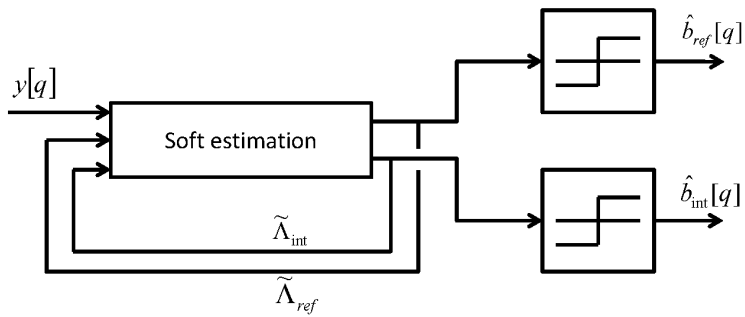

Figure 4. Iterative SIC scheme.

American Institute of Aeronautics and Astronautics

been modified to output soft estimates of de considered signal. The detection/estimation is iterated a number $N_{iter}$ of times, after which a decision is taken on the symbols of both signals. The same method can easily be extended to the dual polarization case especially in case of high cross polarization rejection.

## V. Numerical Results

In this section we present the performance evaluation results for the selected MIMO scheme in two scenarios, namely maritime and land mobile satellite with intermediate tree shadowing (LMS-ITS). The BGAN standard, operating in L-band has been adopted a reference for the physical layer.

We consider the maritime scenario first. The channel model is described in the following table.

| Fast fading | Rician |
|---|---|
| $\mathcal{K}$ Rician factor | 10 dB |
| Doppler shift | 2 Hz |
| # taps | 1 |

**Table 1. Channel parameters**

We focus on bearers types with symbol rate 33.6 Ksymbols/s. More specifically, we focus on the BGAN F80T1Q4B bearer types characterized by QPSK modulation. The code rate for the different F80T1Q4B sub-bearers is described in the following table.

**Table 2. Code Rate of F80T1Q4B bearer types**

| Bearer Name | Coding Rate | Data Rate (Kbps) |
|---|---|---|
| F80T1Q4B-L8 | 0.34 | 21.6 |
| F80T1Q4B-L7 | 0.40 | 25.6 |
| F80T1Q4B-L6 | 0.48 | 30.4 |
| F80T1Q4B-L5 | 0.55 | 35.2 |
| F80T1Q4B-L4 | 0.63 | 40.0 |
| F80T1Q4B-L3 | 0.70 | 44.8 |
| F80T1Q4B-L2 | 0.77 | 49.2 |
| F80T1Q4B-L1 | 0.83 | 52.8 |
| F80T1Q4B-R | 0.87 | 55.6 |

Depending on the geographical location of the user terminal (center/edge of coverage, center/edge of beam) the $C/I$ may vary significantly due to co-channel interference. In order to take this into account we first derived the noise value from the $C/N$ expression:

$$\frac{C}{N} = \frac{PAG}{LKBT} \tag{12}$$

where $P$ is the radiated power, $B$ is the bandwidth, $G$ is the antenna gain at the receiver, $K$ is the Boltzmann constant, $T$ is the antenna noise temperature at the receiver, $L$ is the path-loss and $A$ is the array factor at the transmitter. Hence, we can derive the noise power as:

$$N = BN_0 = \frac{KB}{G/T} \tag{13}$$

From the BGAN standard and common user terminal parameters, we have $B = 200$ KHz, $G/T = 12.5$ dB and $L = 187.05$ dB. Thus the noise power is $N = -133$ dBm. Note that in each simulation the $C/I$ remains constant, as it only depends on the position of the user terminal, while the $C/N$ changes.

As benchmark system we consider one in which a single polarization is considered (SISO system). The same total power per beam $P = \frac{P_{SISO}}{2}$ is assumed in both systems. In order to evaluate the advantage deriving by using both polarizations we use the normalized throughput which is defined as follows

$$\text{Normalized Throughput} = \frac{\text{Throughput Dual Polarization}}{\text{Throughput SISO}}, \tag{14}$$

where

$$\text{Throughput} = (1 - \text{FER})\text{Rate}. \tag{15}$$

In the first simulations we do not assume any correlation between polarizations given that the cross-polar interference is known to be very low in L band. A frequency reuse factor of 4 is assumed. We evaluate the performance of our method for 2 beams representing the best and the worst case scenarios (center and edge of the beam coverage) assuming a user location both at the center of the beam and at the edge. Realistic beam patterns have been considered and co-channel interference from beams at the same frequency as the reference one have been taken into account. We indicate with $C/I$ the ratio between the power of the reference signal and the total interference power. The polarization and beam gain values have been taken from.[19] Perfect channel state information at the receiver (CSIR) is assumed.

Figures 5 and 6 show the throughput of the proposed scheme in the maritime channel normalized to the throughput of the benchmark (SISO) system. It can be seen how, given a code rate, our scheme is able to double the rate at expenses of incrementing the transmit power by 3dB. The impact of the $C/I$ on the system is remarkable since, when considering the beam at the end of coverage, the power increment needed to get twice the SISO throughput is increased up to 4-5 dB. We also notice how bearers with higher channel code rate do not succeed in achieving 100% gain. In order to enhance the throughput for high rate bearers we evaluated the performance of the SIC scheme considered in Section IV for a single polarization setup in a multibeam satellite system with frequency reuse 4 using the same interference pattern as in figure and 6, i.e., edge of beam at edge of coverage. We considered the best case scenario, i.e., AWGN and perfect symbol alignment across all signals. Only the strongest interferer (which has a relative $C/I$ of about 14 dB) is taken into account by the detector/decoder, as the others have much smaller power and would determine an increase in complexity with limited gain in terms of BLER. The results of the simulations are shown in Fig. 7. It can be seen how the SIC gives only a marginal gain even in the best case scenario. This is due to the strong power unbalance between the reference signal and the interferers. As a matter of facts, when the SNR is such that the good signal starts to be de decodable, the stronger interferer is to weak to contribute to the decoding and thus almost no difference is observed in BLER. Other simulations we carried out (not reported here for a matter of space) showed that, in case the $C/I$ relative to the strong interferer and the reference signal have comparable power, SIC can significantly enhance system's BLER in certain bearers. This is the case, for instance, of systems with more aggressive frequency reuse factors (e.g., 2). Thus we conclude that, for the considered setup, SIC does not bring significant improvements. Hence, in the rest of the paper we will not consider SIC techniques.

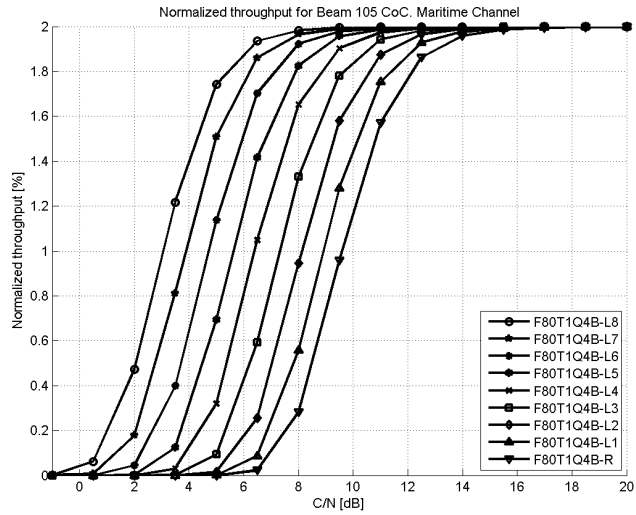

Figure 5. Normalized throughput versus transmit power for polarization-multiplexing scheme and Soft-MMSE-SIC receiver at center of coverage.

In the following we present the results we obtained in the ITS scenario and in a mixed LMS environment (MIX scenario). For such scenario, far more challenging than the maritime one, we used real channel measurements obtained for ITS and MIX scenarios in the context of ESA MIMOSA project. Note that

7 of 11American Institute of Aeronautics and Astronautics

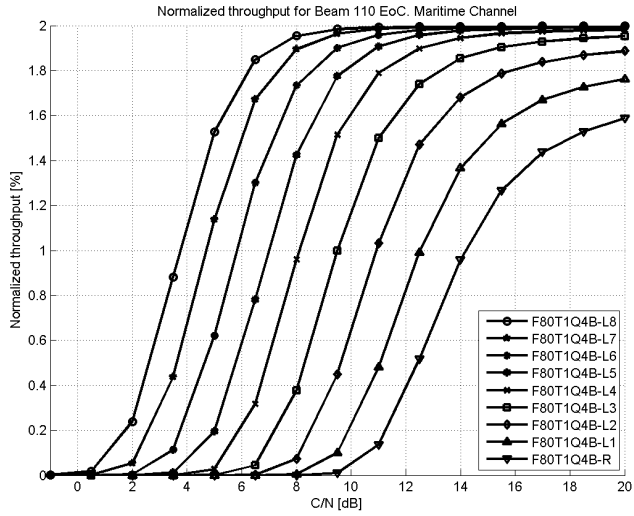

Figure 6. Normalized throughput versus transmit power for polarization-multiplexing scheme and Soft-MMSE-SIC receiver at edge of coverage.

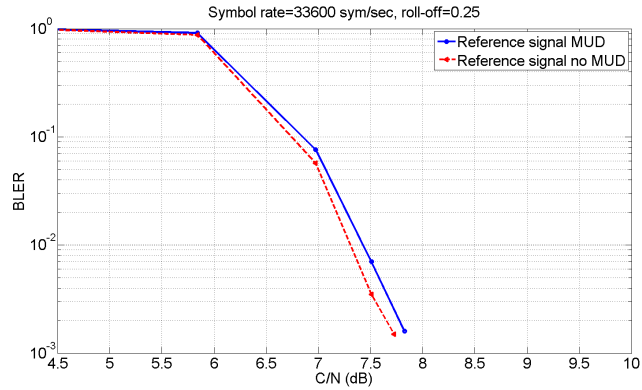

Figure 7. BLER versus $C/N$ in AWGN for the case in which SIC is applied on the reference signal and the strongest interferer in a multibeam satellite system with frequency reuse 4. The six strongest interferers have been considered. Three iterations have been used at the estimator. Bearer F80T1Q4B-L1 with rate 0.825 and QPSK modulation have been used for all signals.

correlation effects are implicitly taken into account in the measurements.

In Fig. 8 part of the measured channel realization in ITS is shown. The alternation of periods of moderate and deep fading can be observed.

Figures 9 and 10 show the normalized throughput for the ITS and MIX channels. It can be seen that a certain throughput gain with respect to the SISO case can be achieved even for the considered channels if the loss in terms of $C/N$ is tolerable. Note also that no interleaver is included in the considered bearers [a]. The inclusion of a time interleaver is likely to enhance the performance of the system significantly and allow to exploit the full potential of dual polarization transmission.

The following observations can be made:

- Given a QPSK modulation with a determined code rate, the use of polarization multiplexing jointly with a Soft-MMSE-SIC demodulator is able to double the data rate at expenses of increasing the transmit power of 3 dB in the best case (in terms of $C/I$) and $4-5$ dB in the worse situations for the maritime scenario.

- When considering more challenging scenarios (ITS,MIX), the use of dual polarization does not provide

---

[a] the BGAN standard includes the use of a time interleaver with 80 msec depth for some higher order bearers

American Institute of Aeronautics and Astronautics

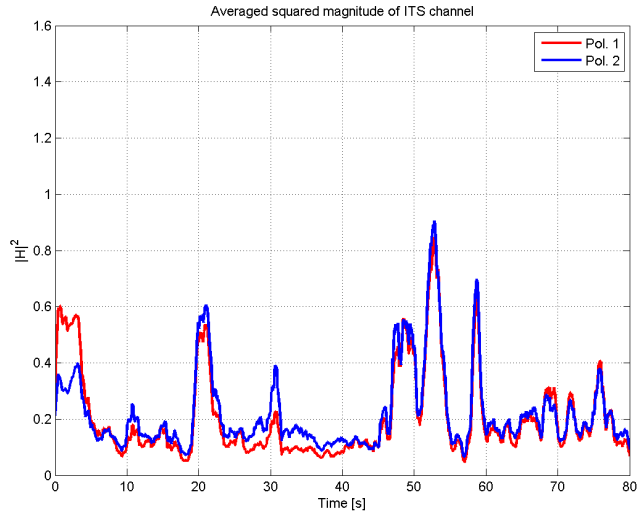

Figure 8. ITS channel realization from real measurements. $h_{11}$ and $h_{22}$ components are shown.

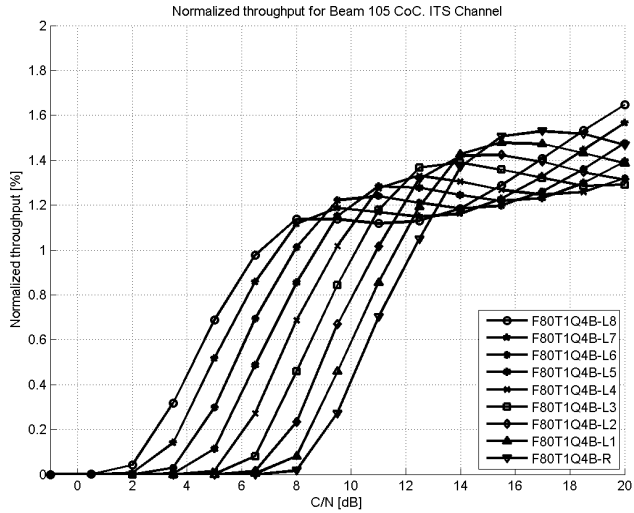

Figure 9. Normalized throughput versus transmit power for polarization-multiplexing scheme and Soft-MMSE-SIC receiver at center of coverage for ITS channel.

the expected gain. This is mainly due to the absence of a time interleaver.

- The fact that the results have been obtained with a low complexity demodulator makes the application of the considered techniques appealing from a practical perspective.

From a system level perspective it is clear that a rate increase can be obtained also by considering more efficient MODCODS (i.e., higher channel code rates and higher order modulations). We evaluated this numerically and concluded that the use of a dual polarization is the best option for the considered setup since the increase in transmit power to obtain twice the throughput as in the SISO case is the less demanding one for a certain level of BLER.

To illustrate this, we compare different ways to double the throughput. Such increase may be achieved using dual polarization, increasing the coderate and increasing the modulation order. In 11 these techniques are normalized by the baseline scenario. The most remarkable aspect is the fact that the spatial multiplexing with dual polarization is 1 dB powerless to achieve the same rate.

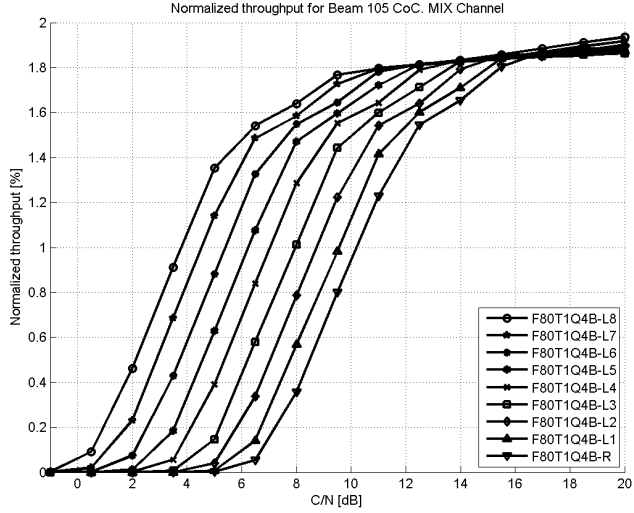

Figure 10. Normalized throughput versus transmit power for polarization-multiplexing scheme and Soft-MMSE-SIC receiver at center of coverage for MIX channel.

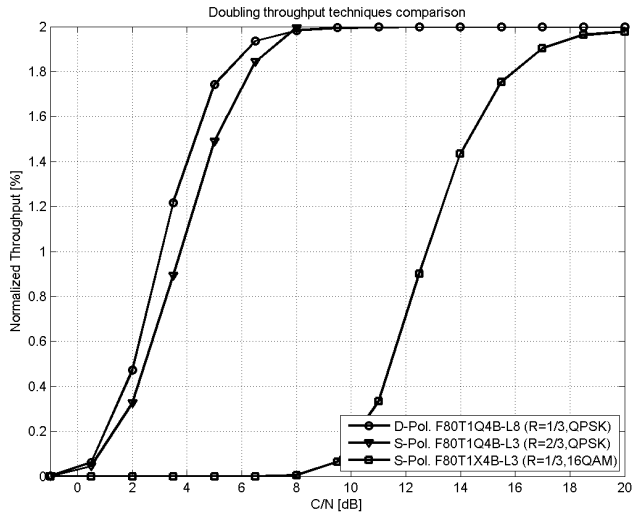

Figure 11. Doubling throughput techniques comparison.

## VI. Conclusions and future work

We presented the results of our study on the application of polarization-time codes and soft interference cancellation in multibeam satellite systems. We adopted the BGAN standard and used realistic channel models and interference patterns.

As future work we plan to extend the simulations to higher order modulations, investigate the use of single user precoding as described in the current high data rate terrestrial standards. As such techniques require feedback from the receiver the impact of system round-trip delay will be also studied. The use of the Soft-MMSE-SIC demodulator must be investigated also for the mobile broadcast standards where the use of long interleavers might increase the performance of the system yet maintaining a low complexity receiver. Another possible research line is to study systems with more aggressive frequency reuse factors using joint SIC and MIMO techniques, that are likely to provide good results in terms of throughput and system availability.

# Acknowledgements

The present work has been carried out under the ARTES 1 programme founded by the European Space Agency.

The view expressed herein can in no way be taken to reflect the official opinion of the European Space Agency.